\documentclass[12pt]{iopart}

\usepackage{graphicx} 
\usepackage{cite}
\usepackage{ulem}
\usepackage{iopams}  
\begin{document}



\renewcommand{\r}{({\bf r})}
\newcommand{\rp}{({\bf r'})}

\newcommand{\ua}{\uparrow}
\newcommand{\da}{\downarrow}
\newcommand{\la}{\langle}
\newcommand{\ra}{\rangle}
\newcommand{\dg}{\dagger}

\title[Energy parametrizations for the 1D Hubbard model: gaps, spin and inhomogeneity]{Simple parametrization for the ground-state energy of the infinite Hubbard chain incorporating Mott physics, spin-dependent phenomena and spatial inhomogeneity}

\author{Vivian V. Fran\c{c}a$^{1,2}$, Daniel Vieira$^3$ and Klaus Capelle$^4$}

\address{1. Physikalisches Institut, Albert Ludwigs Universit\"at, Hermann-Herder 3, 79104, Freiburg, Germany\\2. Capes Foundation, Ministry of Education of Brazil, Caixa Postal 250, Brasilia, 70040-20, Brazil\\3. Departamento de F\' isica,
Universidade do Estado de Santa Catarina, Joinville, 89219-710 SC, Brazil\\4. Centro de Ci\^encias Naturais e Humanas, Universidade Federal do ABC (UFABC), Santo Andr\'e, 09210-170 SP, Brazil}


\begin{abstract}
Simple analytical parametrizations for the ground-state energy of the one-dimensional repulsive Hubbard model are developed. The charge-dependence of the energy is parametrized using exact results extracted from the Bethe-Ansatz. The resulting parametrization is shown to be in better agreement with highly precise data obtained from fully numerical solution of the Bethe-Ansatz equations than previous expressions [Lima et al., Phys. Rev. Lett. {\bf 90}, 146402 (2003)]. Unlike these earlier proposals, the present parametrization correctly predicts a positive Mott gap at half filling for any $U>0$. The construction is extended to spin-dependent phenomena by parametrizing the magnetization-dependence of the ground-state energy using further exact results and numerical benchmarking. Lastly, the parametrizations developed for the spatially uniform model are extended by means of a simple local-density-type approximation to spatially inhomogeneous models, e.g., in the presence of impurities, external fields or trapping potentials. Results are shown to be in excellent agreement with independent many-body calculations, at a fraction of the computational cost.
\end{abstract}

\maketitle

\section{Introduction}
\label{intro}

The archetypical strong-correlation phenomenon is the Mott insulator \cite{mottbook}. It is well known that the first-principles description of such systems by means of density-functional theory (DFT) encounters severe difficulties. The single-particle (Kohn-Sham) gap calculated with standard DFT methodology is not the same as the many-body gap, even in principle and if no approximations at all are made during the calculation. In the particular case of the Mott insulator, it is known that a proper description of the Mott gap is obtained by adding to the single-particle Kohn-Sham gap a correction arising from the derivative discontinuity of the exchange-correlation functional \cite{pplb,mcy1,mcy2,mcy3,rubio1,rubio2}.

Many modern density functionals do have an implicit derivative discontinuity due to their orbital dependence, but this affords at best an incomplete description of the Mott state and the Mott gap \cite{mcy1,rubio1,rubio2}. Only very few density functionals have an explicit discontinuity as a function of the density, among them the 2-electron functional of Mori-Sanchez, Cohen and Yang \cite{mcy1} which is based on the so-called flat-plane condition, devised by the same authors,\cite{mcy1,mcy2,mcy3} and the Bethe-Ansatz LDA (BALDA) of Lima et al. \cite{lsoc,mottEPL}, which is based on an approximate analytical parametrization of the Bethe-Ansatz solution for the ground-state energy of the one-dimensional Hubbard model that becomes exact in several important limits. Both of these functionals do properly account for the Mott insulator. Only the BALDA, however, has been parametrized in a way that allows its application to a wide range of physical systems. As a consequence, BALDA and variations thereof has been applied to inhomogeneous correlated many-electron systems as well as to correlated many-atom systems in optical lattices and traps \cite{mottEPL,kurthPRL,mirjaniPRB,sanvito,schenk,verdozzi,progress,superlatticePRB,rapcom,franca}. 

However, in these applications it has become clear that the parametrization by Lima et al., also referred to as LSOC parametrization, after the initials of its developers, still does not provide a fully correct description of the Mott gap. In particular, at small U, the LSOC expression for $E_c$ has a derivative discontinuity, but the resulting Mott gap is negative, effectively predicting the Mott state to be a (strange) metal instead of an insulator. It is not trivial to correct this behaviour, as any change to the LSOC expression must preserve the exact limits and properties already built into it. Thus, instead of merely algebraic adjustments, it becomes necessary to understand and improve the physics missing from the LSOC parametrization in this regime. 

A key aspect of the Mott insulator is that it is a nonmagnetic state of matter, i.e., the insulating nature is not the result of antiferromagnetism. Therefore, the solution to the problem just described must be expressed, within DFT, in terms of the charge-density only, and cannot make use of spin densities and spin-density-functional theory (SDFT). On the other hand, many correlated systems do have magnetic phases. Therefore, a more complete description of strong correlations must properly account for both, the insulating state and various types of magnetic states, as well as their possible coexistence. While such description is possible within Bethe-Ansatz based SDFT by performing a fully numerical solution of the spin-resolved Bethe-Ansatz equations and interpolating between the resulting data points every time the exchange-correlation energy needs to be evaluated, this procedure is very inconvenient for Kohn-Sham calculations, where the functional must be evaluated hundreds or thousands of times during the iterations towards selfconsistency. A simple parametrization of the spin dependence would allow straightforward application of BA-DFT methodology to spin-dependent phenomena in electronic systems and hyperfine-label-dependent phenomena in optical lattices.  

The present paper reports progress along both of these lines. In Section \ref{mottsec} we identify a shortcoming of the standard parametrization used in BALDA and propose a simple, {\it ad hoc} but physically motivated, variation of it that is more accurate and whose Mott gap properly remains positive for all values of the interaction $U$. The revised parametrization also considerably improves the description of the metallic (e.g., Luttinger liquid) phases. In Section \ref{spinsec} we employ exact analytical results extracted from the Bethe Ansatz to further generalize this revised parametrization to spin-dependent situations. In Section \ref{inhomsec} we use a simple local-density type approximation to extend our results to spatially nonuniform systems. Density-Matrix Renormalization Group (DMRG) and Lanczos calculations are performed to test and validate this approximation. Section \ref{summary} contains a brief summary.

\section{Improved description of the Mott gap}
\label{mottsec}

The task at hand is to obtain a simple and accurate analytical approximation for the ground-state energy $E_0$ of the inhomogeneous one-dimensional Hubbard model (1DHM), 
\begin{equation}
\hat{H} = -t \sum_{i, \sigma} \left(\hat{c}_{i \sigma}^{\dagger}\hat{c}_{i+1, \sigma} + {\rm H.c.} \right) 
+ U\sum_{i}  \hat{n}_{i \uparrow}\hat{n}_{i \downarrow} + \sum_{i, \sigma} V_i \hat{n}_{i \sigma},
\end{equation}
where $\hat{n}_{i\sigma}=\hat{c}_{i \sigma}^{\dagger}\hat{c}_{i \sigma}$ is the spin-resolved particle-density operator at site $i$, $\hat{c}_{i\sigma}^\dagger$ and $\hat{c}_{i\sigma}$ are fermionic creation and annihilation operators, $t$ is the hopping parameter (taken to be the unit of energy), $U$ the on-site interaction and $V_i$ an on-site potential that makes the system spatially inhomogeneous. In applications to electrons in crystal lattices, $V_i$ can describe inequivalent atoms in the lattice, while in applications to cold atoms in optical lattices it accounts for the trapping potential.

The commonly used expression for the per-site ground-state energy ($e_0$) of the 1DHM is the LSOC parametrization, given by \cite{lsoc}
\begin{equation}
e_0^{LSOC}(n,U)=
-\frac{2 \beta(U)}{\pi} \sin \left(\frac{\pi n}{\beta(U)}\right),
\label{lsocparam}
\end{equation}
where $n=n_\uparrow+n_\downarrow$ is the charge density and the interaction $U$ enters $e_0(n,U)$ through the interaction function $\beta(U)$, which is determined from
\begin{equation}
-\frac{2\beta(U)}{\pi}\sin\left(\frac{\pi}{\beta(U)}\right)=-4\int_0^\infty dx \frac{J_0(x)J_1(x)}{x\left(1+e^{Ux/2}\right)}.
\label{halffilling}
\end{equation}
By construction, this expression becomes exact for $U\to 0$ and any $n$ (where $\beta=2$), for $U\to \infty$ and any n (where $\beta=1$), and for $n=1$ and any $U$ (where $0 \leq \beta \leq 1$), and provides a reasonable approximation to the full Bethe-Ansatz solution inbetween. 

In the LSOC parametrization, the interaction function $\beta(U)$ is independent of the particle density and of spin. This independence is algebraically very convenient, as it allows one to determine $\beta(U)$ outside the selfconsistency cycle of DFT instead of having to recalculate it any time the charge (or spin) density changes. However, it is physically incorrect, as the relation between the bare interaction parameter $U$ and the correlation energy $E_c$ must depend on the charge density, {\it e.g.} through screening. As a consequence of the density-independence of $\beta$, LSOC retains the sinusoidal density-dependence of the energy, which is correct only at $U=0$ and $U\to\infty$, also for all intermediate values of $U$. 

\begin{figure}[htb]
\centering
\includegraphics[width=11cm]{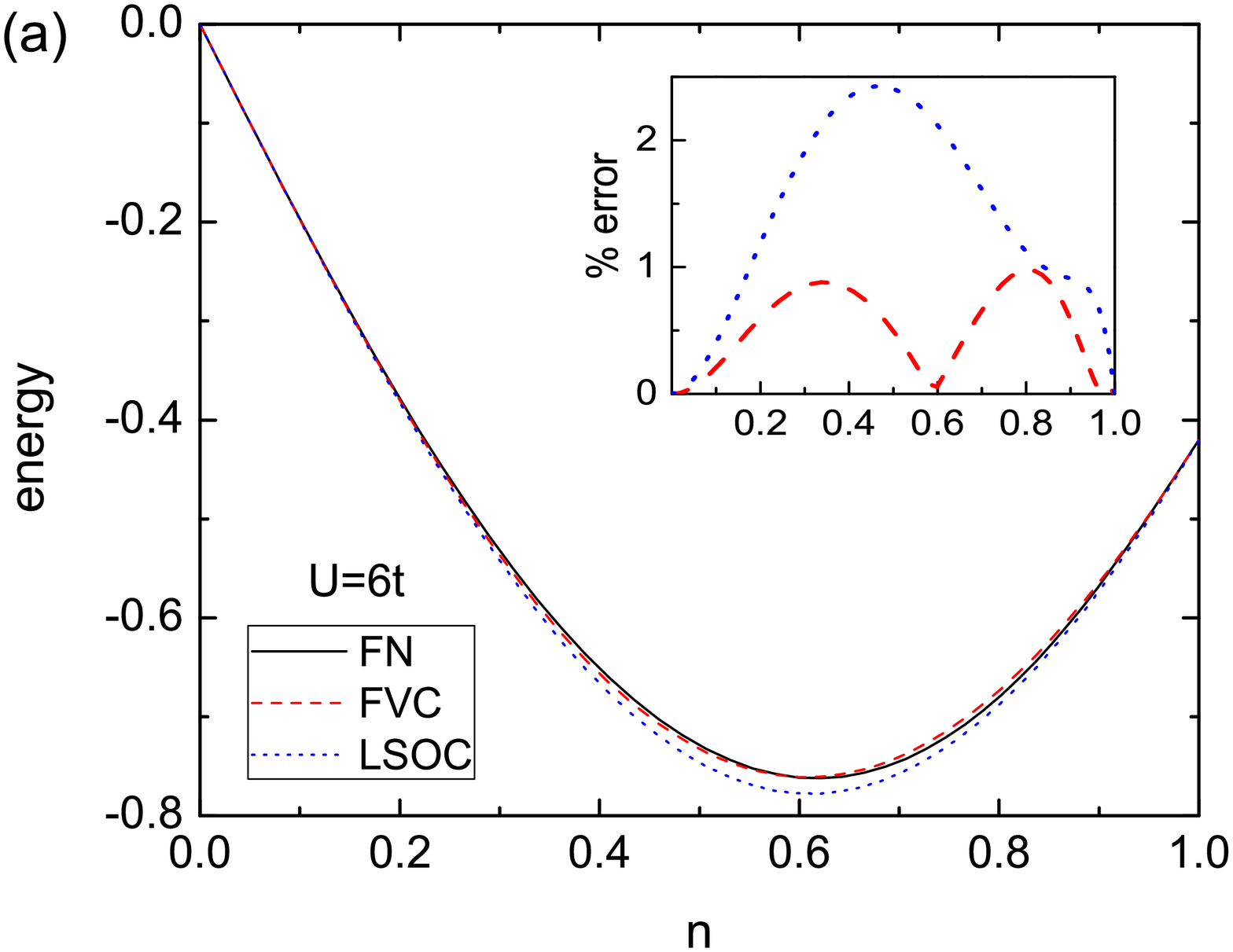}\vspace{-0.2cm}
\includegraphics[width=11cm]{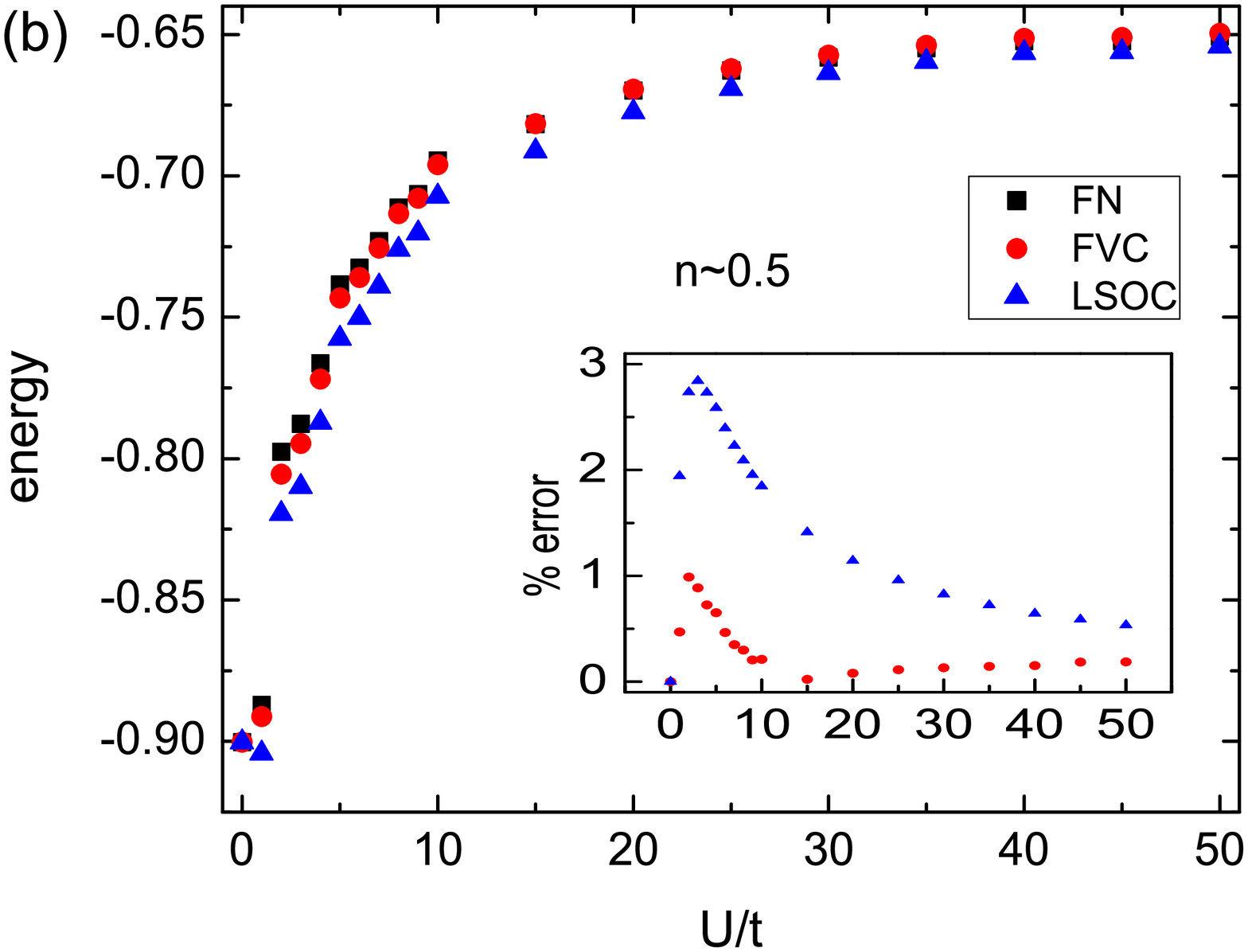}
\caption{Per-site ground-state energy as a function of: a) density, for $U=6t$; b) interaction, for $n\sim 0.5$. Insets: percentage deviation from the fully numerical (FN) data, defined as $100(e^{FVC(LSOC)}-e^{FN})/e^{FN}$.}
\label{fig1}
\end{figure}  

Thus, it becomes necessary to modify the interaction function in a way that allows it to change with the density. The additional density dependence, however {\it must not spoil the behaviour in the three limits exactly recovered by LSOC}. This condition severely restricts the modifications that could possibly be made to the LSOC parametrization. The form we adopt is
\begin{eqnarray}
e_0^{FVC}(n,U)=
-\frac{2 \beta(n,U)}{\pi} \sin \left(\frac{\pi n}{\beta(n,U)}\right)
\label{fvcnospin}
\end{eqnarray}
where $\beta(n,U)=\beta(U)^{\alpha(n,U)}$ and $\alpha(n,U)=n^{\sqrt[3]{U}/8}$. We stress that this particular form has not been derived from first principles but is a physically motivated {\it ad hoc} modification designed to restore the density-dependence of the interaction function through the replacement $\beta(U)\to \beta(n,U)$, while preserving all exact limits obeyed by the LSOC expression. The specific form chosen for the exponent $\alpha(n,U)$ is a consequence of the later generalization to spin-dependent phenomena, as explained in Sec.~\ref{spinsec}.
   
Figure~\ref{fig1} compares the present parametrization (\ref{fvcnospin}) to data obtained from a fully numerical (FN) solution of the Bethe-Ansatz integral equations. For comparison  purposes, the earlier LSOC parametrization (\ref{lsocparam}) is also included. To distinguish the present from the LSOC expression, we label the curves corresponding to the former by FVC. As the insets show, the relative deviation of FVC data from FN data is typically less than $2\%$, and at most $\sim 4\%$. This is the same size of error of the local-density approximation itself (quantified by comparing BALDA/FN data for small Hubbard chains to results from exact diagonalization), so that to within the accuracy of the LDA the present parametrization is a faithful representation of the full BA solution for the entire parameter range, including values of $U \gg 6t$ that cannot be realized in solids but occur in systems of trapped cold atoms.
 
\begin{figure}
\centering
\includegraphics[width=11cm]{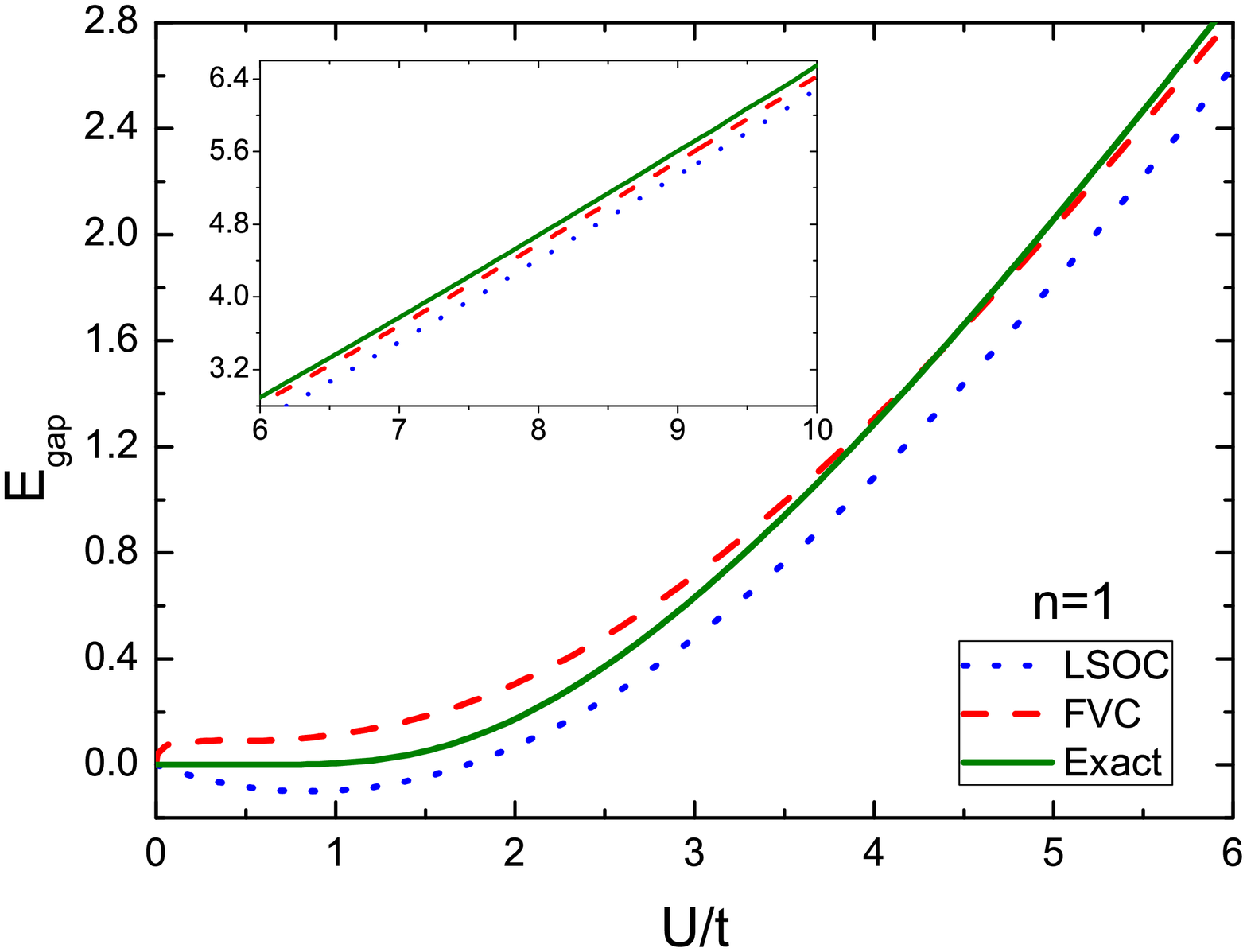}
\caption{Mott gap $E_{gap}$ of the infinite homogeneous chain obtained from numerical solution of the full BA equations, compared to analytical predictions obtained from the LSOC and the present (FVC) parametrization.}
\label{fig2}
\end{figure}

Our main motivation for developing Eq.~(\ref{fvcnospin}), however, was the incorrectly negative Mott gap predicted by the LSOC expression between $U=0$ and $U=2t$. The Mott gap $E_{gap}$ can be evaluated analytically from the expression for $e_0(n,U)$, either by explicitly calculating the derivative discontinuity or by taking total-energy differences \cite{mottEPL}. From the LSOC parametrization one obtains \cite{mottEPL}
\begin{eqnarray}
\label{egaplsoc}
E^{LSOC} _{gap} (U)&=& U + 4 \cos\left(\frac{\pi}{\beta(U)}\right),
\end{eqnarray}
whereas the present parametrization leads to
\begin{eqnarray}
\label{egapfvc}
E^{FVC}_{gap} (U) &=& U + 4 \cos\left(\frac{\pi}{\beta(U)}\right)- \frac{1}{2}\beta(U) U^{1/3} \ln \beta(U) \times\nonumber\\
&&
\hspace{-0.17cm}\left[-\frac{1}{\pi}\sin\left(\frac{\pi}{\beta(U)}\right) + \frac{1}{\beta(U)}\cos\left(\frac{\pi}{\beta(U)}\right)\right].
\end{eqnarray}
As illustrated in Fig.~\ref{fig2}, the additional terms arising from the present expression push the gap upward, avoiding that it becomes negative between $U=0$ and $U=2t$ as was, incorrectly, the case within the LSOC expression \cite{mottEPL}. For $U>2t$ both the LSOC and the present (FVC) gap are positive, the latter being significantly closer to the numerical BA results than the former, although neither reproduces the subtle nonperturbative behavior for $U\to 0$, in spite of the logarithmic term in Eq.~(\ref{egapfvc}). Only within the latter, however, the Mott gap is everywhere positive.

\section{Extension to spin-dependent phenomena}
\label{spinsec}

In spin-polarized situations, the energy depends on the spin density $m=n_\uparrow - n_\downarrow$, in addition to the charge density $n=n_\uparrow + n_\downarrow$ and the interaction $U$. This dependence is not included in the LSOC parametrization, which can therefore not be applied to study spin-dependent phenomena for electrons in solids or hyperfine-polarization-dependent phenomena for atoms in optical lattices. We note that BALSDA calculations for the 1DHM have already been performed in the context of cold atoms in optical lattices \cite{vivaldoPRB,danielJMMM,hu}. In lieu of an analytical parametrization, these works resorted to a fully numerical (FN) solution of the BA integral equations. An analytical parametrization would substantially simplify this approach, making the BALSDA as easily implementable as the BALDA one. 

An additional advantage of the analytical approach over the numerical one is that in the solution of the BA integral equations one cannot specify from the outset the density and magnetization of the system one is interested in. Rather, one has to specify the upper and lower limits of the integrals, and obtains the densities as part of the solution. This is clearly inconvenient for DFT, where 
the energies have to be evaluated as functions of the densities. More generally, it is desirable to be able to specify the system under study in terms of physical observables, such as the densities $n$ and $m$, instead of in terms of auxiliary quantities, as the limits of the BA integrals. 

Analytical expressions also permit one to derive further analytical results for other quantities. Our present derivation of closed expressions for the Mott gap is one example, and the analytical derivation and solution of Euler equations determining the phase diagram of harmonically trapped fermions on optical lattices \cite{rapcom} is another. For all these reasons, a simple but reliable parametrization of $e_0(n,m,U$) can be useful for various types of calculations, within SDFT and beyond. Therefore, we next present an analytical parametrization of $e_0(n,m,U)$ and use it to construct a Bethe-Ansatz local-spin-density approximation (BALSDA) for the 1DHM. 

In order to generalize the LSOC and FVC expressions to spin-dependent situations, we one more time follow the basic philosophy of constructing an analytical interpolation function recovering exactly known limits as function of the physically relevant variables. Several such exact results for $e_0(n,m,U)$ are known \cite{lw1,lw2,schlottmann}. For non-interacting systems ($U=0$), 
\begin{equation}
e_0(n,m,U=0)=-\frac{4}{\pi}\sin\left(\frac{\pi n}{2}\right)\cos\left(\frac{\pi m}{2}\right)\label{limit1}.
\end{equation}
For infinite interaction ($U\rightarrow \infty$),
\begin{equation}
e_0(n,m,U\rightarrow\infty)=-\frac{2}{\pi}\sin(\pi n)\label{limit2}.
\end{equation}
For half-filled unpolarized systems $(n=1, m=0)$,
\begin{equation}
e_0(n=1,m=0,U)=-4\int_0^\infty dx \frac{J_0(x)J_1(x)}{x\left(1+\exp^{Ux/2}\right)}.\label{limit4}
\end{equation}
Finally, for maximum magnetization ($m=n$),
\begin{equation}
e_0(n,m=n,U)=-\frac{2}{\pi}\sin(\pi n).
\label{maxspin}
\end{equation}

The LSOC and FVC parametrizations take $m=0$, and express $e(n,U)$ as a controlled analytical interpolation between (\ref{limit2}), (\ref{limit4}) and the $m=0$ limit of (\ref{limit1}). The construction of a more complete interpolation, recovering all four limits as functions of $n$, $m$ and $U$, is strongly constrained by these limits, but still not unique. We therefore impose five additional common-sense criteria: (i) avoid high-order polynomials, which can produce unphysical wiggles, (ii) avoid unusual special functions, and (iii) keep the form similar to the LSOC parametrization. This third condition is useful because by now the LSOC parametrization has been implemented and used by many groups \cite{mottEPL,kurthPRL,mirjaniPRB,sanvito,schenk,verdozzi}, so it will be easier to update to a new parametrization which has a similar form to the old one. However, as we have argued above, the LSOC expression does not properly describe the density-dependence of the ground-state energy at intermediate $U$. Therefore, we also require, as condition (iv), our spin-dependent generalization to reduce to the present Eq.~(\ref{fvcnospin}) for $m=0$ instead of to Eq.~(\ref{lsocparam}). The final, fifth, additional condition is motivated by computational efficiency and is explained in Sec.~\ref{inhomsec} below. Even with these additional common-sense criteria, the form of the parametrization is not uniquely determined, and a very large function space can still be explored. The particular choice made below was obtained by starting from the LSOC form and then building in, one-by-one, the additional exact limits and criteria. Many different variations have been explored, but we stopped when arriving at one whose deviation from the fully numerical solution of the Bethe-Ansatz equations was less than the typical error bar of the local-density approximation for this type of system. At this point, further improvements in the form of the parametrization become indistinguishable, in applications to inhomogeneous systems, from the intrinsic error of the LDA.

All four exact limits and five supplementary conditions are incorporated by the choice
\begin{eqnarray}
e_0^{FVC}(n,m,U)=
-\frac{2 \beta(n,m,U)}{\pi} \sin \left(\frac{\pi n}{\beta(n,m,U)}\right) \cos \left(\frac{\pi m}{\gamma(n,m,U)}\right),
\label{fvcparam}
\end{eqnarray}
where
\begin{equation}
\beta(n, m, U) = \beta(U)^{\alpha(n,m,U)},\label{b(n,m,U)}
\end{equation}
\begin{equation}
\gamma(n,m,U) = 2 \exp\left[\frac{\sqrt{U}}{1-(m/n)^{3/2}}\right],
\end{equation}
and
\begin{equation}
\alpha(n, m, U) = \left[\frac{n^2 - m^2}{n^{15/8}}\right]^{\sqrt[3]{U}}.
\label{alphadef}
\end{equation}
Here, $\beta(U)$ is the same quantity employed in the LSOC parametrization. For zero magnetization, $\alpha(n,m,U)$ reduces to $\alpha(n,U)$ used in Eq.~(\ref{fvcnospin}). Equation (\ref{fvcparam}) is valid for $U\geq 0$ and $n\leq 1$, but it can be extended to $n>1$ and to $U<0$ by standard particle-hole transformations \cite{lw1,schlottmann,mottEPL,rapcom}.

\begin{figure}[h]
\centering
\includegraphics[width=10cm]{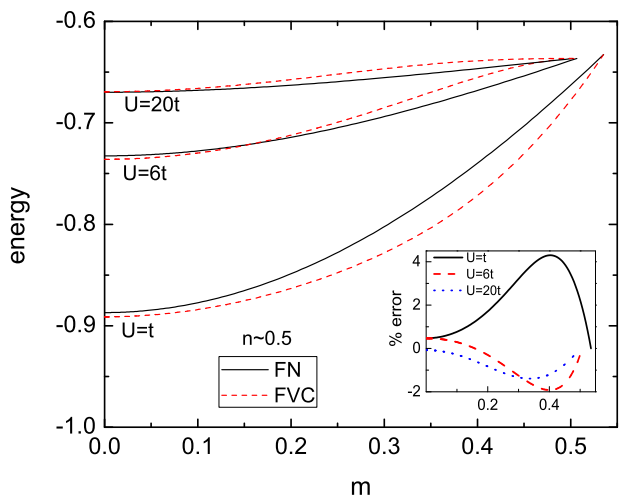}\vspace{-0.5cm}
\caption{Per-site ground-state energy as a function of magnetization, for $n\sim 0.5$ and several values of $U$. Inset: percentage deviation from the fully numerical (FN) data, defined as $100(e^{FVC(LSOC)}-e^{FN})/e^{FN}$.}
\label{fig3}
\end{figure} 

Satisfaction of condition (\ref{maxspin}) is of particular interest because this condition has no counterpart in the spin-independent situation. Moreover, it establishes a coupling between the charge and the spin dependence. Physically, maximum spin means that the Pauli principle keeps all fermions maximally apart for any $U$, just as infinitely repulsive interactions ($U\to \infty$) do for any degree of spin-polarization. For $n=m$ the spin-dependent parametrization must thus reduce to the same limit as for $U\to\infty$ and recover the value $\beta=1$. On the other hand, for $m=0$ the expression should reduce to the earlier form (\ref{fvcnospin}). This double requirement is the explanation of the particular form chosen for $\alpha(n,m,U)$ of Eq.~(\ref{alphadef}) and, consequently, for $\alpha(n,U)$ of Eq.~(\ref{fvcnospin}).

Figure \ref{fig3} presents a comparison of the spin-dependence of this parametrization with data obtained from a fully numerical solution of the BA integral equations for intermediate parameter values, where the expression is not exact already by construction. Clearly, the spin-dependence of the homogeneous system is recovered to within the same precision as the charge dependence. 

\section{Local approximation for inhomogeneous systems}
\label{inhomsec}

The main use of DFT and SDFT is in calculations for spatially inhomogeneous systems, where translational symmetry is broken and the density becomes position dependent. In the case of lattice models, inhomogeneity means that not all sites are equivalent. By means of the L(S)DA prescription, expressions for the energy of the homogeneous system, such as those described in the preceding section, can be used on a site-by-site basis to approximate the corresponding energy of the  inhomogeneous system.

Explicitly, the local-spin-density approximation to any energy component $E$ is given by
\begin{eqnarray}
\label{lda}
E \approx E^{LSDA} = \sum_i^L e(n,m,U)|_{^{n \rightarrow n_i}_{m \rightarrow m_i}},
\end{eqnarray}
where $L$ is the number of lattice sites, and  $e = \lim_{L\to \infty}E(L)/L$ is the per-site energy of the homogeneous system. This expression approximates the energy of the inhomogeneous system, where $n_i$ and $m_i$ vary from site to site, by evaluating the energy density of the homogeneous system site by site at the densities of the inhomogeneous one. In DFT, including model DFT, this prescription is usually applied to the correlation energy, which for the Hubbard model can be defined as $E_c= E_0 - E_{MF}$, where $E_{MF}$ is the mean-field approximation to the ground-state energy $E_0$. Since $E_{MF}$ is simple to obtain, the task to approximate the correlation energy $E_c[n_i,m_i,U]$ of 
the inhomogeneous Hubbard model, is thus reduced to that of approximating the per-site ground-state energy of the homogeneous one, $e_0(n,m,U)$. This is the quantity that we extracted above from the Bethe-Ansatz equations.

The minimization of the resulting energy functional is conveniently carried out via selfconsistent Kohn-Sham (KS) calculations of the charge and spin densities and of the resulting ground-state energies. In such KS calculations the correlation potentials (obtained by differentiating the correlation energy with respect to $n$ and $m$, or, equivalently, to $n_\uparrow$ and $n_\downarrow$) are evaluated once in each of the iterations of the selfconsistency
cycle. In these iterations, the densities $n_i$ and $m_i$ change, but the interaction $U$, being a parameter of the Hamiltonian, does not. We have therefore expressed the integral in Eq.(\ref{halffilling}) and the resulting transcendental equation for $\beta$ exclusively in terms of $U$, in order to guarantee that the only slightly time-consuming steps of the calculation take place only once in each calculation, outside the selfconsistency cycle. This is the fifth additional condition on the spin-dependent parametrization, alluded to above.

\begin{figure}[!t]
\centering
\includegraphics[width=17cm]{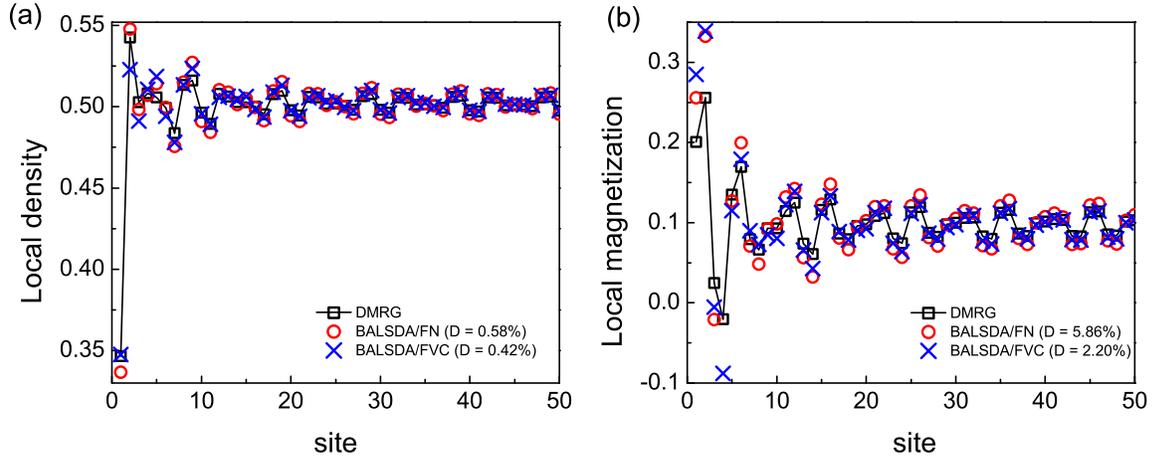}
\vspace{-6.5cm}
\caption{(a) Density profile $n_i$ of an open chain with $L=100$ sites, $N_\uparrow=30$, $N_\downarrow=20$ fermions, and $U=4t$, obtained selfconsistently from fully numerical BALSDA, the present parametrization and DMRG calculations. (b) Local magnetization $m_i$ of the same system. Percentage deviations, summed over all sites, of the LSDA densities from the DMRG ones are given in parenthesis. }
\label{fig4}
\end{figure}

As a simple example of such KS calculations, which serves to illustrate all essential aspects, we consider open Hubbard chains, where the spatial inhomogeneity stems from the boundaries, which give rise to charge and spin-density oscillations in the bulk. Representative results for a chain with $L=100$ sites are displayed in Fig.~\ref{fig4}, where we compare the ground-state density profile (Fig.\ref{fig4}--a) and spin-density profile (Fig.\ref{fig4}--b) obtained from DMRG to LSDA profiles obtained from using the fully numerical solution to the BA integral equations and from the present parametrization. The BALSDA/FVC and BALSDA/FN ground-state energies of the same system deviate from the DMRG energy by $0.01\%$ and $0.64\%$, respectively. The local densities follow the same trend, and deviate from DMRG by $0.42\%$ and $0.58\%$. For the local magnetization, the corresponding numbers are $2.20\%$ for BALSDA/FVC and $5.86\%$ for BALSDA/FN. Remarkably, for all three quantities the parametrized results are closer to the DMRG benchmark data than the numerically defined BALDA. This shows that the particular form chosen for the proposed parametrizations allows for considerable error cancellation. On 32 processors the DMRG calculation took approximately 17 hours, while the BALSDA calculations required approximately 40 seconds. Of course, for high-precison calculations, as well as for the calculation of quantities that are not easily extracted from densities and energies, DMRG is still essential. 

\begin{figure}[!t]
\centering
\includegraphics[width=15.7cm]{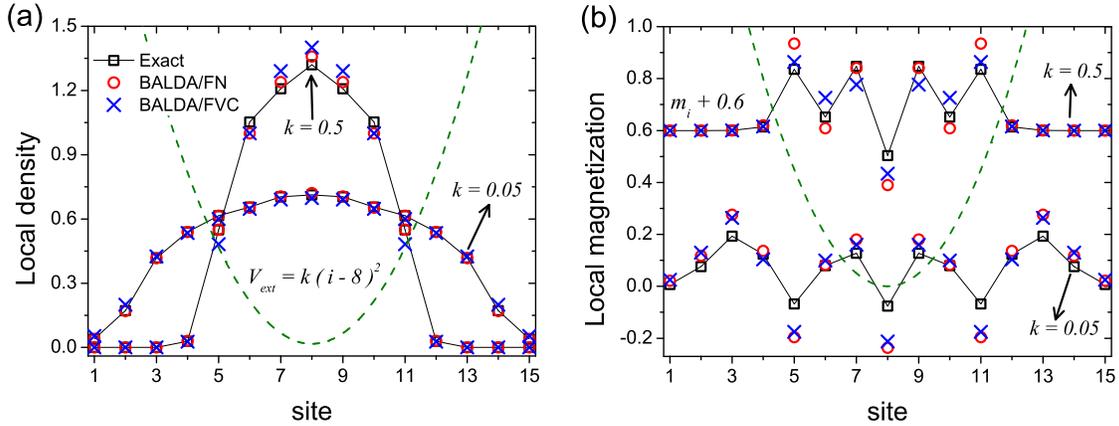}
\caption{Parabolically confined systems: (a) Density profiles $n_i$ of periodic chains for two different values of curvature ($k=0.05$ and $k=0.5$). $L=15$ sites, $N_\uparrow=4$, $N_\downarrow=3$ fermions, and $U=4t$, obtained selfconsistently from fully numerical BALSDA, the present parametrization and exact (Lanczos) calculations. (b) Local magnetization $m_i$ of the same systems. }
\label{fig5}
\end{figure}

A more complex case is depicted in Fig.~\ref{fig5}, which shows density and magnetization profiles for parabolically confined systems in a periodic chain. For two different values ($k=0.05$ and $k=0.5$) of the curvature of the confining potential (whose form is schematically indicated by the dashed (green) curve), the data points show the site-resolved particle density and spin density obtained by exact (Lanczcos) diagonalization, fully numerical BA-LSDA and our presently proposed parametrization. To within the accuracy that can be expected from a local-density approximation for this type of system (a few percent) the agreement between all three sets of calculations is excelent, for both charge and spin distributions. The amplitude of the magnetization-density oscillations is overestimated by both flavours of local approximations, which is consistent with previous observations for similar approximations and systems \cite{danielJMMM}.

\begin{figure}[!t]
\centering
\includegraphics[width=10cm]{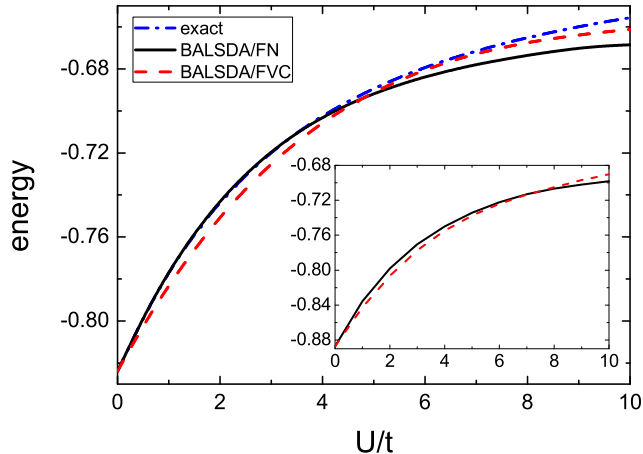}
\caption{Per-site ground-state energy of finite open chains with $L=15$ sites, $N_\uparrow=4$ and $N_\downarrow=3$ fermions, obtained selfconsistently from fully numerical BALSDA, the present parametrization and from exact diagonalization. Inset: BALSDA data for a larger chain with $L=200$ sites, $N_\uparrow=60$ and $N_\downarrow=40$.}
\label{fig6}
\end{figure}

Next, we compare, in Fig.~\ref{fig6}, the numerically exact ground-state energy obtained from Lanczos diagonalization of a small open Hubbard chain with $L=15$ sites to the LSDA ground-state energies obtained from using the fully numerical solution to the BA integral equations and from using the present parametrization. Up to $U\sim 4t$ the BALSDA/FN data are almost identical to the exact data, which attests to the quality of the local approximation. For $U$ larger than $\sim 5t$, the present parametrization is, once again, better than the conceptually superior fully numerical LSDA, due to error cancellation. The inset shows that for larger systems the behaviour is qualitatively the same.

Finally, we point out that successful applications of our (then unpublished) spin-dependent expression (\ref{fvcparam}) to the study of spin-polarized transport across a correlated nanoconstriction \cite{mirjaniPRB} and to the calculation of occupation probabilities of exotic superfluids in spin-imbalanced systems \cite{abu} have already been reported.

\section{Summary}
\label{summary}

In summary, we have constructed a simple and reliable parametrization for the ground-state energy of the homogeneous one-dimensional Hubbard model, for arbitrary fillings, spin-polarizations and interactions. For the first time, a qualitatively and quantitatively correct description of the Mott gap is obtained from a simple density functional with a proper explicit derivative discontinuity. This parametrization can be used in its own right, for the homogeneous model, whenever simple expressions for the ground-state energy and for the resulting Mott gap are required. 

However, its main application is as input for local-density and local-spin-density approximations, which allow one to efficiently minimize the energy and extract energies, density profiles and related quantities for spatially  inhomogeneous models. Since in KS calculations one never diagonalizes the interacting Hamiltonian, but only the auxiliary noninteracting one, systems with thousands of sites can be dealt with, even in the absence of any simplifying symmetry. 

This work was supported by Brazilian agencies CAPES, CNPq and FAPESP. The authors thank Dominik H\"orndlein for the DMRG data and Vivaldo Campo for his code for solving the Bethe-Ansatz integral equations and the numerically defined BA-L(S)DA.

\end{document}